\documentclass[aps,pra,twocolumn,groupedaddress,nofootinbib, nobibnotes]{revtex4-1}
\usepackage{epsfig,amssymb,amsmath,amsfonts}
\usepackage{times}
\usepackage{epstopdf}
\usepackage{graphicx}
\usepackage{dcolumn}
\usepackage{bm}
\usepackage{amsthm}
\usepackage{float}
\usepackage{braket}
\usepackage{caption}
\usepackage{subcaption}
\usepackage{tabularx}
\usepackage{color,array}
\usepackage{xcolor}

\def\>{\rangle}
\def\<{\langle}

\newcommand{\ignore}[1]{}

\begin{document}

\title{Fault Tolerance with Bare Ancillary Qubits for a $[[7,1,3]]$ Code}

\author{Muyuan Li}
\author{Mauricio Guti\'errez}
\author{Stanley E. David}
\author{Alonzo Hernandez}
\author{Kenneth R. Brown}
\email{ken.brown@chemistry.gatech.edu}
\affiliation{Schools of Computational Science and Engineering, Chemistry and Biochemistry, and Physics, \\Georgia Institute of Technology, Atlanta, Georgia 30332-0400, USA}


\begin{abstract}

We present a $[[7, 1, 3]]$ quantum error-correcting code that is able to achieve fault-tolerant syndrome measurement using one ancillary qubit per stabilizer for an error model of independent single-qubit Pauli errors. All single-qubit Pauli errors on the ancillary qubits propagate to form exclusively correctable errors on the data qubits. The situation changes for error models with two-qubit Pauli errors. We compare the level-1 logical error rates under two noise models: the standard Pauli symmetric depolarizing error model and an anisotropic error model. The anisotropic model is motivated by control errors on two-qubit gates commonly applied to trapped ion qubits. We find that one ancillary qubit per syndrome measurement is sufficient for fault-tolerance for the anisotropic error, but is not sufficient for the standard depolarizing errors. We then show how to achieve fault tolerance for the standard depolarizing errors by adding flag qubits to check for errors on select ancilary qubits.  Our results on this $[[7, 1, 3]]$ code demonstrates how physically motivated noise models may simplify fault-tolerent protocols.

\end{abstract}
\pacs{}
\keywords{quantum error correction; quantum computing}

\maketitle

\section{Introduction}\label{Sec:Intro}

The operation of a universal quantum computer requires protocols to protect it from instrumental and environmental imperfections. Quantum error-correcting (QEC) \cite{shor1995scheme} codes are among the most promising approaches to deliver scalable and reliable fault-tolerant quantum computation. Topological QEC codes have a two-dimensional qubit layout that limits qubit interactions to only nearest neighbors, which is suitable for practical implementation using quantum architectures such as superconducting devices and trapped ions \cite{raussendorf2007fault, fowler2009high, yoder2016surface}. Concatenated codes require more resources to be mapped onto local architectures, but can use less resources overall than topological codes for low error rate systems \cite{svore2006noise}. In general, the optimal choice of quantum error-correcting code is dependent on the actual noise environment and the quantum circuit being used \cite{suchara2013comparing}. Many QEC codes, both topological codes and concatenated codes, belong to the family of Calderbank-Shor-Steane (CSS) codes, where $X$ and $Z$ Pauli errors can be treated separately \cite{steane1996multiple, calderbank1996good}.

For a given QEC code, different methods of syndrome measurement, state preparation and decoding can all affect the resulting logical error rate.  For syndrome measurement, using a single (bare) ancillary qubit to measure a stabilizer is in general not fault-tolerant, since errors on the ancillary qubit can propagate to the data and form uncorrectable errors \cite{nielsen2010quantum, preskill1997FTQC}.  Although there are various methods to make the syndrome measurement fault-tolerant, these come at the expense of extra resources.  Shor's method requires a $w$-qubit cat state to measure a weight-$w$ stabilizer \cite{shor1996catstate}.  An extra qubit is needed to verify the ancillary qubit, but this is not a strict requirement \cite{aliferis2007noverif, tomita2013comparison}.  Steane's method requires the fault-tolerant preparation of a logical state \cite{steane1997FTQC}, while Knill's method relies on the fault-tolerant preparation of a logical Bell pair \cite{knill2005noisydevices}.

Using bare ancillary qubits for stabilizer measurement can be fault-tolerant if we are guaranteed that single-qubit errors or errors that occur with a probability linearly proportional to the physical error rate do not propagate to form uncorrectable errors.  More specifically, if we assume only Pauli errors, then after measuring a stabilizer of weight $w$ a weight $\lfloor w/2 \rfloor$, where $\lfloor x \rfloor$ is the floor of $x$, error can propagate to the data.  This can still be fault-tolerant in several cases.  In some codes, the presence of gauge subsystems allow the decomposition of high weight stabilizers into lower weight gauge operators that can be measured in a fault-tolerant fashion using bare ancillary qubits.  This is the case of the Bacon-Shor codes \cite{bacon2006BScodes, aliferis2007subsystem}.  In other cases, codes have a large enough distance such that any error of weight up to $\lfloor w/2 \rfloor$ can be corrected ($d \geq 2 \lfloor w/2 \rfloor + 1$).  This is the case of large distance surface \cite{fowler2009high, fowler2012surface} and color codes \cite{bombin2006colorcode, landahl2011color}. Finally, certain codes with distance $d < 2 \lfloor w/2 \rfloor + 1$ allow for stabilizer measurement with bare ancillary qubits because the resulting errors on data qubits are correctable. This is the case for the $[[9,1,3]]$ surface code, where the weight-$2$ errors that propagate when measuring the weight-$4$ stabilizers are all correctable for specific orderings of the entangling two-qubit gates \cite{tomita2014low}. 

In this paper, we present a non-CSS $[[7,1,3]]$ QEC code that falls into the latter category: every single-qubit Pauli error on the ancillary qubit propagates to the data qubits to form a correctable error.  The code was found by a numerical greedy search for stabilizer codes where single-qubit errors on the ancillary qubits do not lead to a logical error \cite{private}.  We refer to this code as the Bare $[[7,1,3]]$ code, and we refer to the syndrome measurement method of one ancillary qubit per stabilizer as the bare method.

Given the guarantee that when performing the bare method any single-qubit Pauli error on the ancillary qubits would not lead to a logical error, we want to understand the behavior of the code under higher weight errors. Here we present the results of Monte Carlo stabilizer simulations of the Bare $[[7,1,3]]$ code with $3$ error correcting steps.  We used CHP \cite{CHP} and a Python-based wrapper to obtain logical error rates for the Bare $[[7,1,3]]$ code at the first level of encoding. We report the logical error rates under two different error models.  We find that with the bare method, fault-tolerance is not achieved under the standard depolarizing error model because, although the code is resilient to any single-qubit error on the ancillary qubit, certain two-qubit errors whose probability is linearly proportional to the physical error rate are malignant. To protect our code against these errors, we present how to use two additional flag qubits \cite{chao2017fault} to achieve fault-tolerant syndrome measurement, which we refer to as the flag method, and report the resulting level-1 pseudothreshold. The bare method is still fault-tolerant under an anisotropic error model applicable to trapped-ion qubits. We report the level-$1$ pseudothreshold of the code under this error model. In the Appendix we present two alternative Monte Carlo error sampling schemes that we employed in the simulations, and we report the level-1 pseudothreshold for two well studied distance-$3$ quantum error correcting codes: the Steane code \cite{steane1996error, gutierrez2015incoherent, gutierrez2016coherent} and the $5$-qubit code \cite{divincenzo1996fault, bennet1996entqecc}, for comparison.

The paper is organized as follows. Section II introduces the Bare $[[7,1,3]]$ code, and explains how it handles different errors. Section III describes the two noise models under which we study the performance of the Bare $[[7,1,3]]$ code and demonstrates why the bare method for this error-correcting code is not fault-tolerant under the standard depolarizing noise model. Section IV describes how to use an additional flag qubit to achieve fault-tolerant syndrome measurement for the Bare $[[7,1,3]]$ code. In Section V, we present the simulation scheme and how the simulation results are used to calculate the level-1 pseudothreshold. Section VI presents the results obtained from our simulation for the Bare $[[7,1,3]]$ code under two error models with two different syndrome measurement methods.
 
\section{Details of the Bare $[[7,1,3]]$ code}\label{sec:code_details} 

The Bare $[[7,1,3]]$ code was found through a numerical greedy search of stabilizer codes with the property that single-qubit errors on the bare ancillary qubit would not lead to a logical error.  Table \ref{table:stabs} presents the stabilizer generators and logical $X$ and $Z$ operators of the code.  As seen from the stabilizers, the code is non-CSS and degenerate.  

\begin{table}[h]
\begin{center}
\begin{tabular}{ c|c }
 \hline
 Stabilizer Generators & Logical Operators \\ \hline \hline
 $X_0 X_4$ & \\
 $X_1 X_4$ & \\
 $X_2 X_5$ & $X_L = X_1 X_2 X_3$\\
 $X_3 X_6$ & $Z_L = Z_0 Z_1 Z_4$\\
 $Z_2 Z_3 Y_5 Y_6$ & \\
 $Z_0 Z_1 Z_2 X_3 Z_4 Z_5$ & \\ \hline
\end{tabular}
\caption{List of stabilizers and logical operators $X_L,\,Z_L$ for the Bare $[[7,1,3]]$ code.}
\label{table:stabs}
\end{center} 
\end{table} 

\begin{figure}[ht]
\centering
\includegraphics[width =0.6\linewidth]{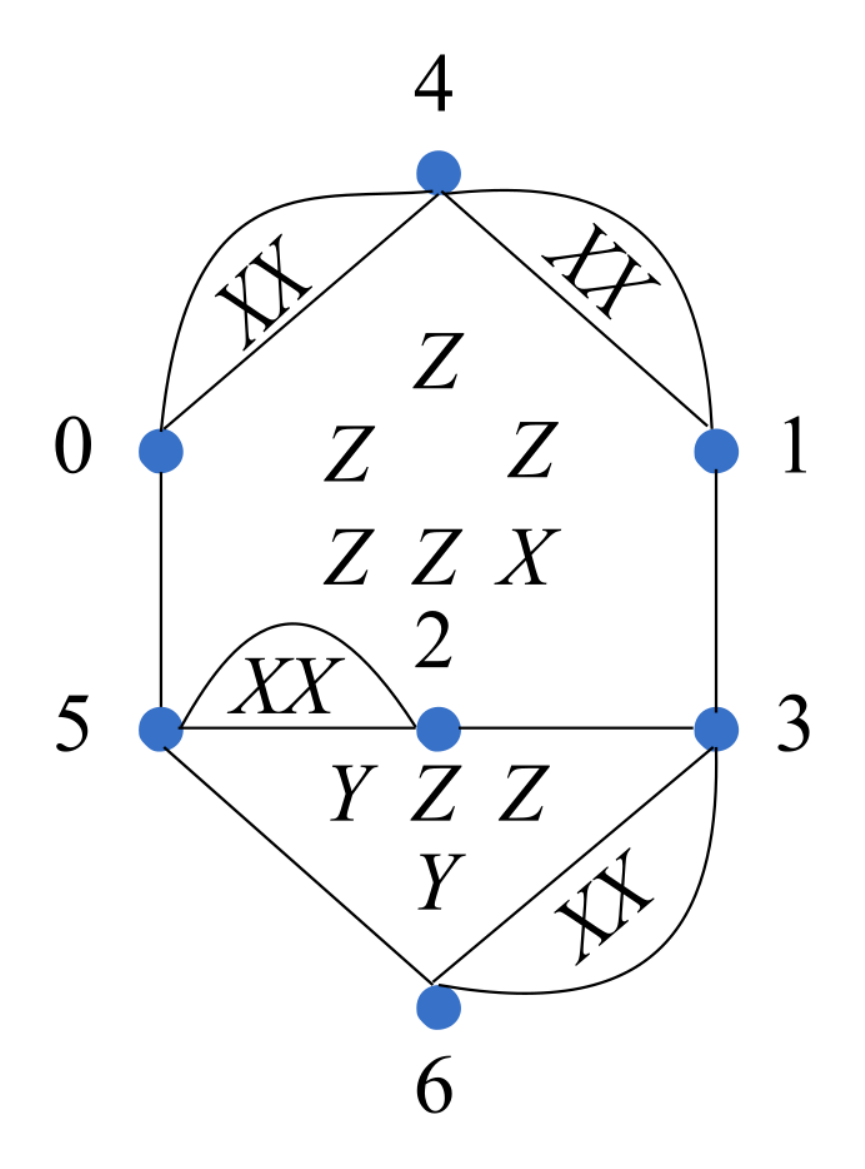}
\captionsetup{justification=raggedright, singlelinecheck=false}
\caption{The Bare $[[7,1,3]]$ code embeded in a plane. Each vertex represents a data qubit, and each stabilizer generator of the code corresponds to a face in this graph.}
\label{fig:crossstabs}
\end{figure}

FIG \ref{fig:crossstabs} shows the configuration of the stabilizers when the Bare $[[7,1,3]]$ code is embeded in a plane.

\subsection{Single-qubit errors on a bare ancillary qubit}

\begin{figure}
\centering
\includegraphics[width =0.9\linewidth]{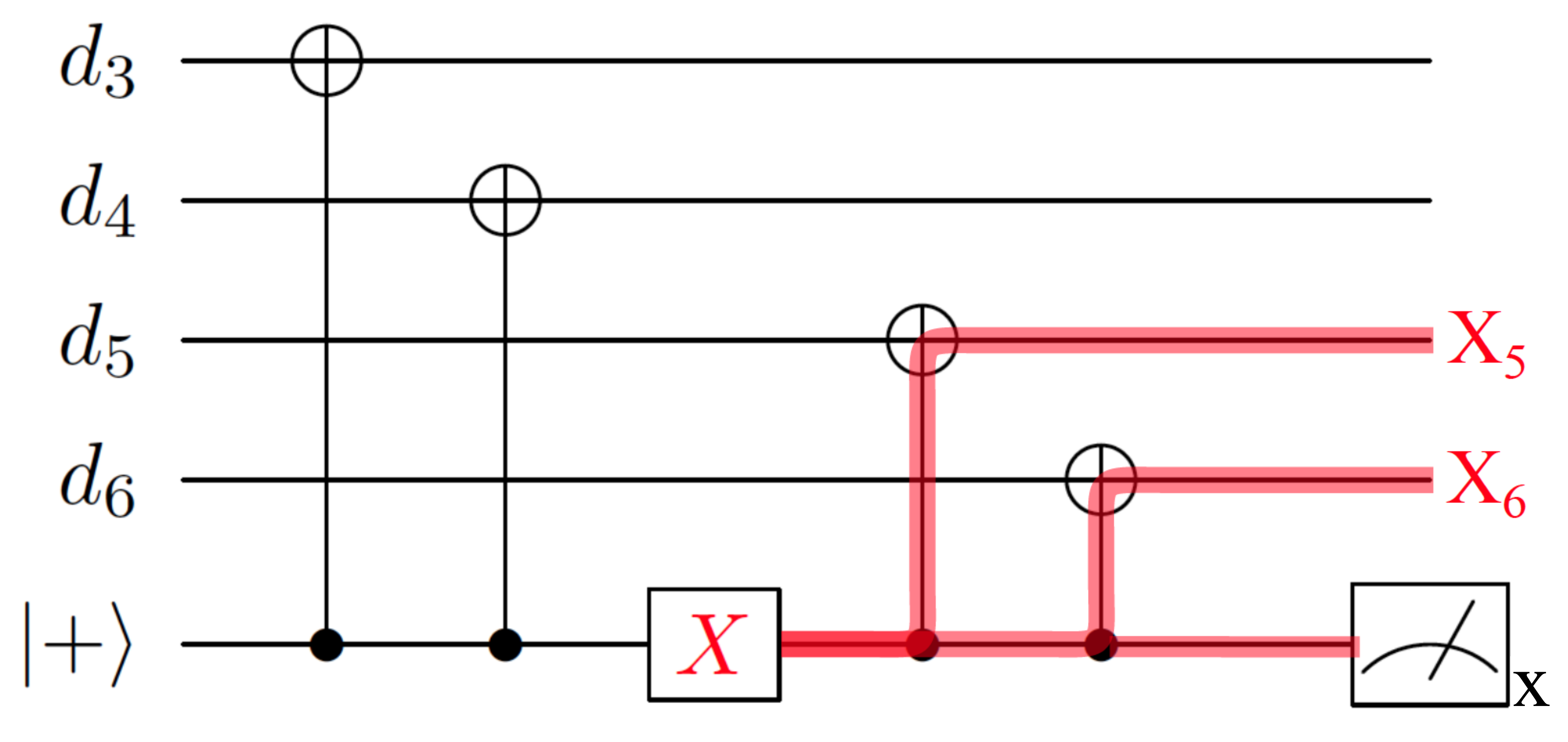}
\captionsetup{justification=raggedright, singlelinecheck=false}
\caption{For the Steane $[[7,1,3]]$ code with bare ancillary qubits, a single-qubit error on the ancillary qubit would lead to a logical error.}
\label{fig:steane}
\end{figure}

To illustrate how using bare ancillary qubits can lead to uncorrectable errors in some QEC codes, let us consider the $[[7,1,3]]$ Steane code.
From Figure \ref{fig:steane}, we can see that when measuring the stabilizer $X_3 X_4 X_5 X_6$, a single qubit $X$ error on the bare ancillary qubit propagates to form the error $X_5 X_6$ on the data qubits.  Although this error is detectable at a later stage, its syndrome is equivalent to an $X_0$ error.  When this correction is applied, the resulting error $X_0 X_5 X_6$ is equivalent to the logical operator $X_L = X_0 X_1 X_2 X_3 X_4 X_5 X_6$ up to a stabilizer.

On the other hand, in the Bare $[[7,1,3]]$ code, all single-qubit errors on the ancillary qubits propagate to become errors with unique syndromes, hence correctable. This is achieved by considering different schedules for coupling the data to the ancillary qubit when measuring the stabilizers. Note that all ancillary qubits are prepared in the $\ket{+}$ state.
\begin{table}[h]
\begin{center}
\begin{tabular}{ c|c|c }
 \hline
 $Z_0 \rightarrow 100000$ & $X_0 \rightarrow 000001$ & $Y_0 \rightarrow 100001$\\
 $Z_1 \rightarrow 010000$ & $X_1 \rightarrow 000001$ & $Y_1 \rightarrow 010001$\\
 $Z_2 \rightarrow 001000$ & $X_2 \rightarrow 000011$ & $Y_2 \rightarrow 001011$\\
 $Z_3 \rightarrow 000101$ & $X_3 \rightarrow 000010$ & $Y_3 \rightarrow 000111$\\
 $Z_4 \rightarrow 110000$ & $X_4 \rightarrow 000001$ & $Y_4 \rightarrow 110001$\\
 $Z_5 \rightarrow 001010$ & $X_5 \rightarrow 000011$ & $Y_5 \rightarrow 001001$\\
 $Z_6 \rightarrow 000110$ & $X_6 \rightarrow 000010$ & $Y_6 \rightarrow 000100$\\ \hline
\end{tabular}
\caption{Single-qubit error syndromes for the Bare $[[7,1,3]]$ code.}
\label{table:single}
\end{center}
\end{table}

The syndromes for the $21$ single-qubit Pauli errors are shown in Table \ref{table:single}.  Notice that the syndromes of $Z_2 Z_3$ is $001101$ which is distinct from all syndromes of single-qubit errors. Similarly, $Z_0 Z_2 \rightarrow 101000$, $Z_0 Z_2 X_3 \rightarrow 101010$, and $Z_4 Z_5 \rightarrow 111010$ all have unique syndromes. Since each gate used in a measurement acts between an ancillary qubit and a data qubit where the ancillary qubit controls a Pauli operator on the data qubit, this observation suggests the following syndrome measurement coupling schedule:
\begin{enumerate}
    \item For each weight-$2$ generator, the measurement gates can be coupled to the ancillary qubit in any order.
    \item For the stabilizer $Z_2 Z_3 Y_5 Y_6$, couple the measurement gates in left-to-right order.
    \item For the stabilizer $Z_0 Z_1 Z_2 X_3 Z_4 Z_5$, couple the measurement gates in order of $Z_0, \, Z_2, \, X_3, \, Z_1,\, Z_4, \, Z_5$.
\end{enumerate}
By using this coupling schedule, for the stabilizer $Z_2 Z_3 Y_5 Y_6$, a single-qubit error on the ancillary qubit can propagate the error $Z_2 Z_3$ onto the data qubits, which has a unique syndrome; for the stabilizer $Z_0 Z_1 Z_2 X_3 Z_4 Z_5$, a single-qubit error on the ancillary qubit can propagate errors $Z_0 Z_2$, $Z_0 Z_2 X_3$, and $Z_4 Z_5$ onto the data qubits, which all have unique syndromes. Therefore, all single-qubit errors on the ancillary qubits propagate to exclusively correctable errors.

\subsection{Two-qubit errors on a bare ancillary qubit}
\begin{figure}[ht]
    \centering
    \begin{subfigure}[b]{\linewidth}
    \includegraphics[width = \linewidth]{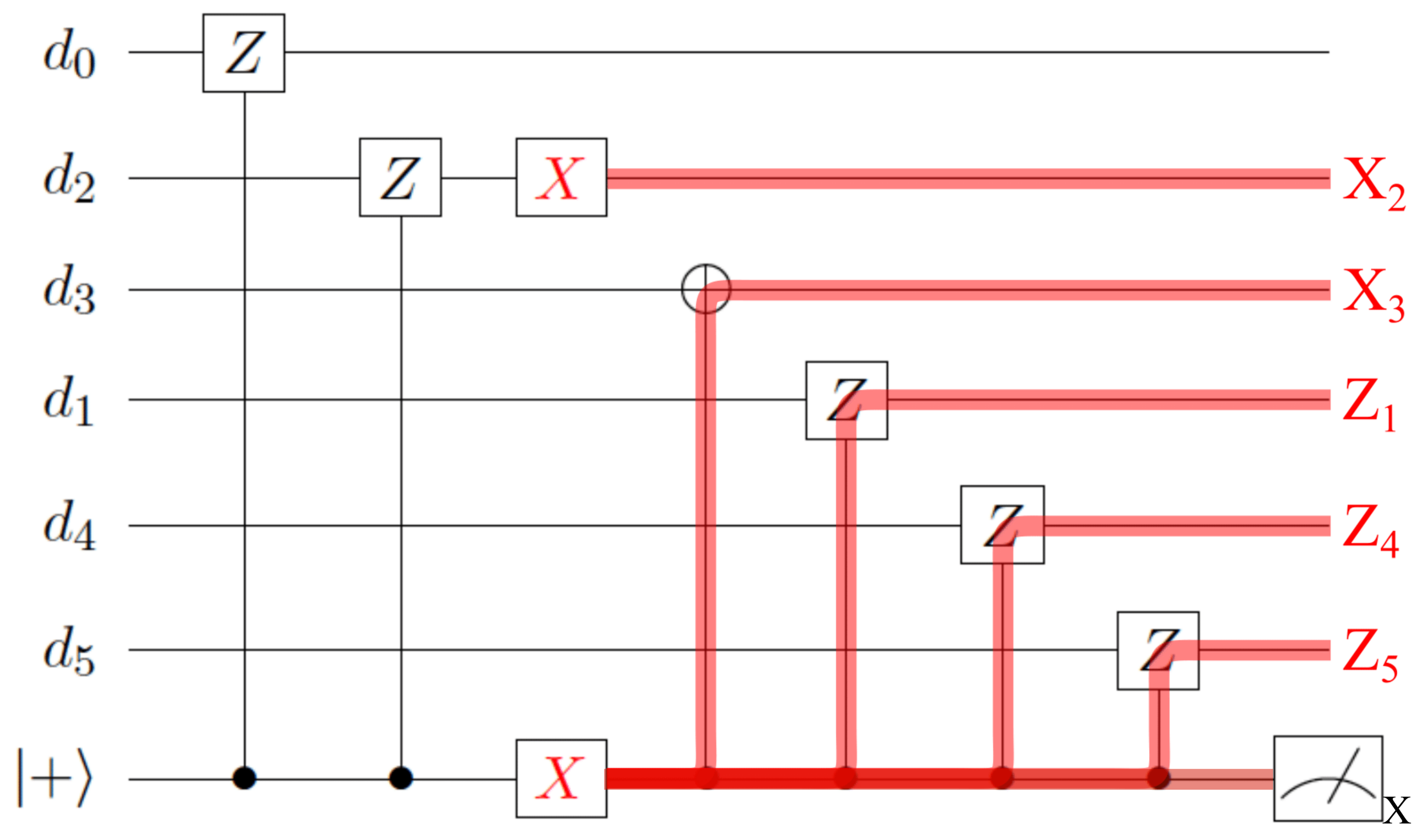}
    \caption{}
    \label{fig:largeweight}
    \end{subfigure}
    ~ 
    \begin{subfigure}[b]{\linewidth}
    \includegraphics[width = \linewidth]{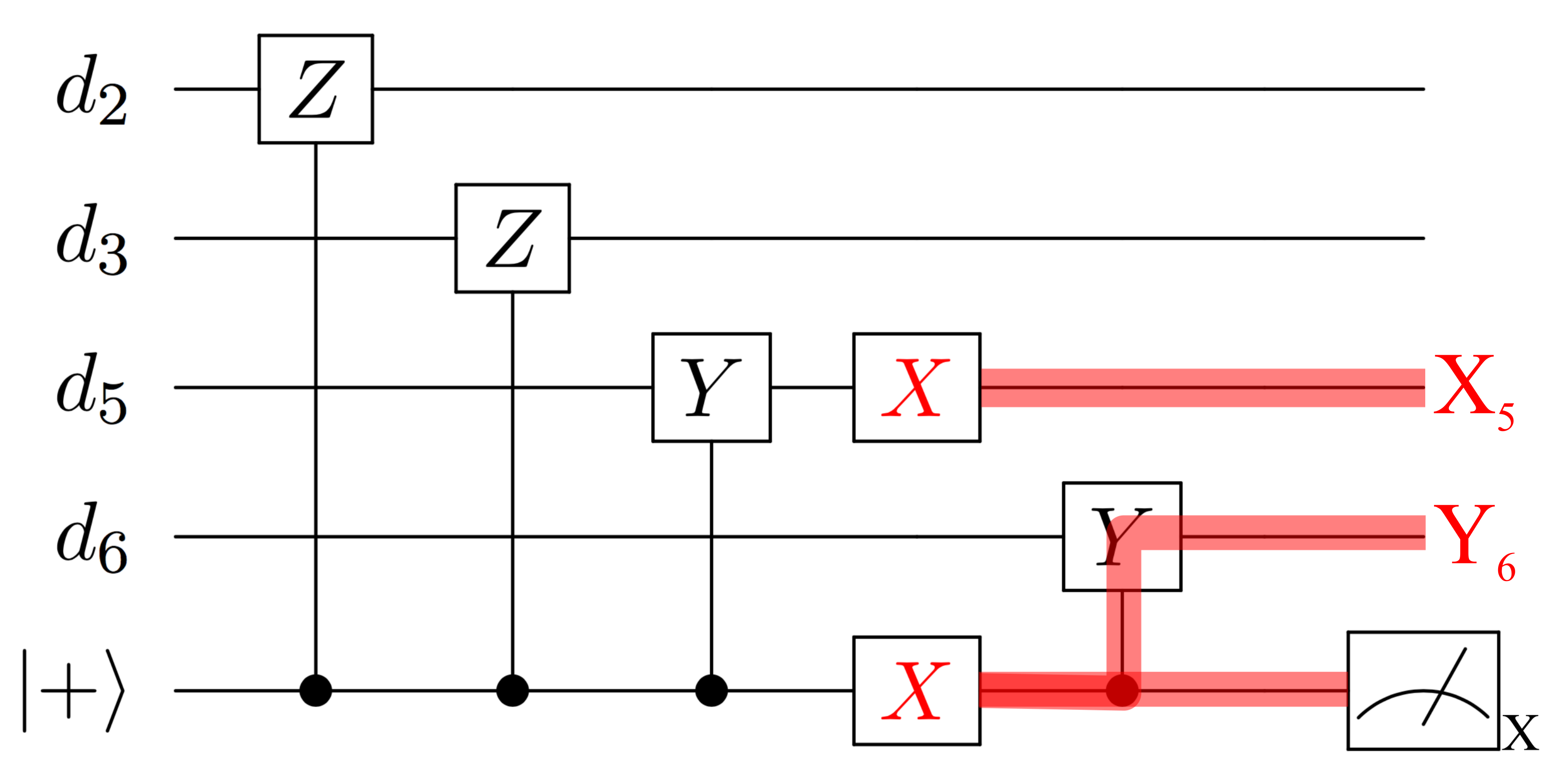}
    \caption{}
    \label{fig:mixture}
    \end{subfigure}
    \captionsetup{justification=raggedright, singlelinecheck=false}
    \caption{Two examples of $2$-qubit $XX$ errors that propagate to become uncorrectable errors.  In (a) the error leads to an $X_L$, while in (b) the error results in a $Z_L$.}
\label{fig:error}
\end{figure}

The Bare $[[7,1,3]]$ code is vulnerable to ``hook errors''.  Figure \ref{fig:largeweight} illustrates an instance of a $2$-qubit Pauli error that can occur under the standard symmetric depolarizing error model.  This $XX$ error propagates to become $Z_1 X_2 X_3 Z_4 Z_5$, which has syndrome $101011$, the same syndrome as $Y_1 Z_4 Z_5$. Thus, the correction $Y_1 Z_4 Z_5$ will be applied, and the resulting error is $X_1 X_2 X_3$, the logical $X$ operator.

In Fig.\ref{fig:mixture} the $2$-qubit $XX$ error on the control-Y gate propagates to become the error $X_5 Y_6$, which has syndrome $000111$, the same as $Y_3$. After the correction $Y_3$ is applied, the resulting error $Y_3 X_5 Y_6$ is equivalent to the logical $Z$ operator up to a stabilizer.

\section{Error Models}\label{sec: error}

\subsection{Error Models}\label{subsec:error}
To study the properties of the Bare $[[7,1,3]]$ code, we applied two Pauli error models: the standard depolarizing error model and an anisotropic error model.

\subsubsection{Standard Depolarizing Error Model}
The standard depolarizing error model is a common quantum channel where, after each gate, there is a symmetric depolarization with some probability $p$.  
After single-qubit gates (including Pauli gates and ancillary qubit preparation gates), a traceless single-qubit Pauli operator, randomly selected from $P=\{X, Y, Z\}$, is applied with probability $p_s$. After two-qubit gates, a traceless two-qubit Pauli operator, randomly selected from $Q=\left \{I,X,Y,Z \right \} \otimes \left \{I,X,Y,Z \right \} \setminus \left \{II \right \}$, is applied with probablity $p_t$. Although in general $p_s \neq p_t$, it is common for them to be simulated with the same value \cite{knill2005noisydevices}. A measurement reports the wrong result with probability $p_{meas}$. For the cases studied here, $p_{meas}$ is set to the single qubit error probability $p_s$. 

\subsubsection{Anisotropic Error Model}

For single qubit gates, preparation, and measurement errors remain the same as the standard depolarizing error model.  The conceptual motivation for the anisotropic error model is that two-qubit gate errors occur due to errors in the gate coupling.  As an example, consider errors in the entangling M{\o}lmer-S{\o}rensen (MS) gate used in trapped ions \cite{MS1999MS}.  The MS gate corresponds to $\exp \left (-i (\pi /4) XX \right )$, and an over or under-rortation could result in the coherent error $\exp \left (-i (\epsilon /2) XX \right )$, where $\epsilon$ denotes the over- or under-rotation angle.  For $\epsilon$ randomly and symmetrically distributed around $0$, this error can be described as a random application of the Pauli error $XX$ with probability $p_t$, which is determined by the distribution of $\theta$.

The key idea is that the two-qubit correlation in the error is not any two-qubit error, but only the two-qubit Pauli error aligned with the gate.  The error after a Control-$P$ gate is then $ZP$ with probability $p_t$. For the specific case of ion traps, this can be derived by including the single qubit rotations necessary to transform the MS gate into a Control-$P$ gate.  In this study, we do not include the individual rotations but work at the level of Control-$P$ gates. We expect random errors on the individual qubits in addition to the control error from the two-qubit gate. To account for the random errors, after every two-qubit gate we apply the two-qubit anisotropic error with probability $p_t$ followed by single qubit errors with probability $p_s$ on the two qubits involved in the gate.

\subsection{Fault-tolerance Dependent on Error Model}
\label{subsec:FTD}

When performing the bare method under the two error models defined in Section \ref{subsec:error}, we can see that all two-qubit errors in the anisotropic error model can be detected and corrected, because a $Z$ error on the control qubit does not propagate to form other errors on the data. However, for certain two-qubit errors in the standard depolarizing error model, such as the $XX$ error as seen in Fig. \ref{fig:largeweight} and Fig. \ref{fig:mixture}, the errors propagate to form uncorrectable hook errors. Therefore, the Bare $[[7,1,3]]$ code with the bare method can achieve fault-tolerance under the anisotropic error model because all errors that occur with probability linear in the error rate of physical operations are correctable; however, the same syndrome measurement method cannot achieve fault-tolerance under the standard depolarizing error model since certain two-qubit errors that occur with probability linear in the physical error rate propagate to become uncorrectable logical errors. 

\section{Fault-tolerance With a Flag Qubit}
\label{sec:flag}
Since the bare method is not fault-tolerant for the Bare $[[7,1,3]]$ code under the standard depolarizing error model, additional resources would be required to ensure fault-tolerance. In this section we propose the method of using two additional flag qubits to perform fault-tolerant syndrome measurement on the Bare $[[7,1,3]]$ code. The idea of the flag method was used by Yoder and Kim \cite{yoder2016surface}, and presented in detail by Chao and Reichardt \cite{chao2017fault}.

Since the source of logical errors for the Bare $[[7,1,3]]$ code are the hook errors propagated from errors on the ancillary qubit, we can use flag qubits to detect errors on the ancillary qubit that can propagate to form these hooks. In order to use the flag method, we need to change the order of the gates to measure the stabilizer $Z_0 Z_1 Z_2 X_3 Z_4 Z_5$ to: $Z_0, X_3, Z_4, Z_2, Z_1, Z_5$. The purpose of this order change is to ensure that all errors that could trigger the flag qubit measurement have distinct syndromes. As we can see in TABLE \ref{table:flag_errors6}, this change of order makes the Bare [[7,1,3]] code vulnerable to certain single-qubit errors that were correctable with the old order. But since these errors all trigger the flag qubit measurement and all result in distinct syndromes, they are still correctable.

\begin{figure}
    \centering
    \begin{subfigure}[b]{0.45\textwidth}
    \includegraphics[width=\linewidth]{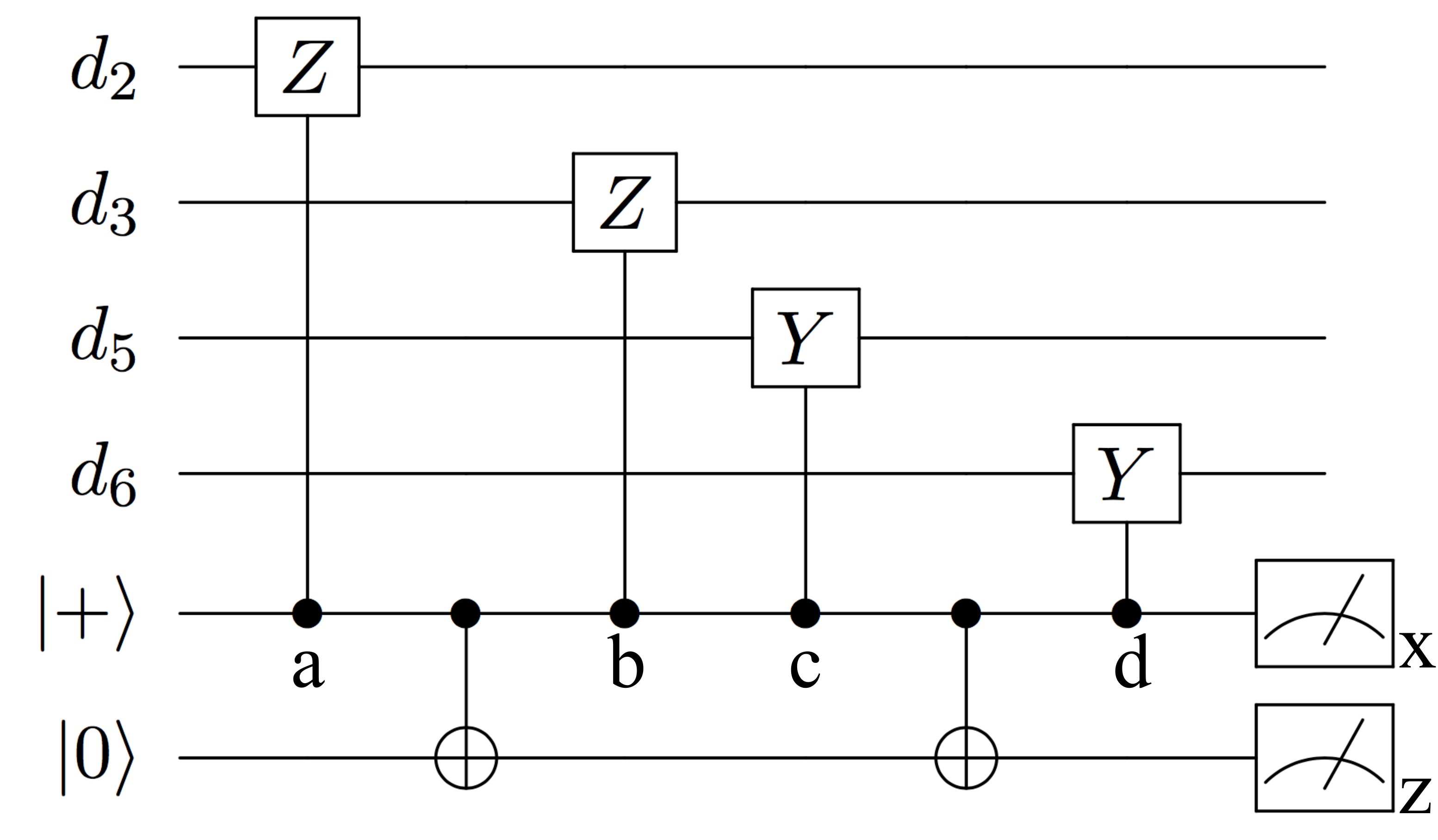}
    \caption{}
    \label{fig:weight4}
    \end{subfigure}
    ~ 
    \begin{subfigure}[b]{0.45\textwidth}
    \includegraphics[width = \linewidth]{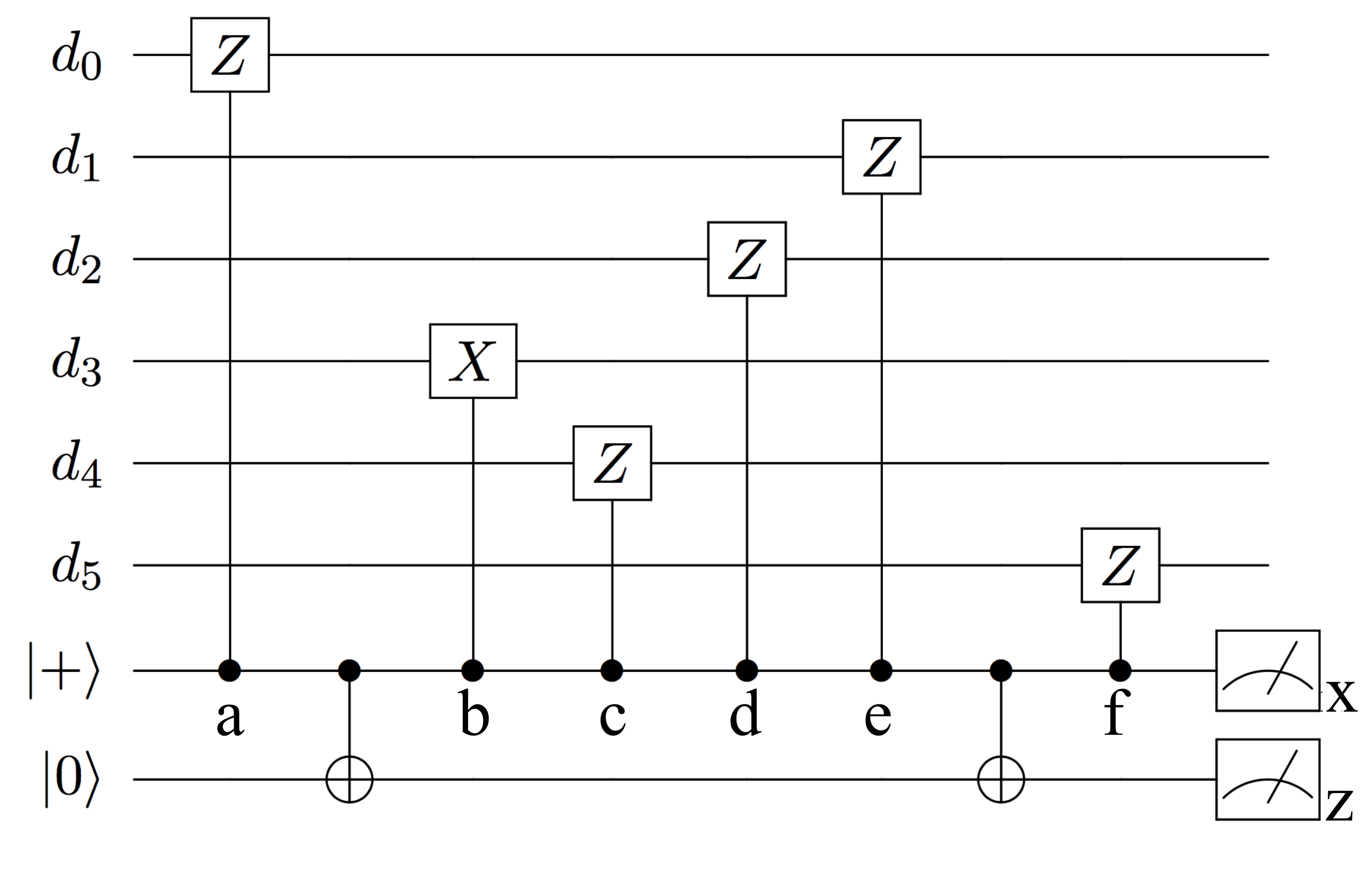}
    \caption{}
    \label{fig:weight6}
    \end{subfigure}
    \captionsetup{justification=raggedright, singlelinecheck=false}
    \caption{Flagged syndrome measurement for the Bare $[[7,1,3]]$ code. (a) Circuit to fault-tolerantly measure the syndrome of the stabilizer $Z_2 Z_3 Y_5 Y_6$ using a flag qubit. (b) Circuit to fault-tolerantly measure the syndrome of the stabilizer $Z_0 Z_1 Z_2 X_3 Z_4 Z_5$ using a flag qubit. Notice the order of gates is different from what we used in the the bare method.}
    \label{fig:flag}
\end{figure}

The circuits to fault-tolerantly measure the high-weight stabilizers are shown in Figure \ref{fig:flag}.  The second ancillary qubit initialized in the $\ket{0}$ state in both circuits is the flag qubit. With no errors present, these circuits behave exactly the same as the circuits using bare ancillary qubits for syndrome measurement, and the $Z$-basis measurements on the flag qubits will always give $\ket{0}$. In Figure \ref{fig:weight4} errors after gate $a$ and $d$ cannot create uncorrectable hook errors, and in Figure \ref{fig:weight6} errors after gate $a$ and $f$ cannot create uncorrectable hook errors. Errors after all other gates in these two circuits will be detected by causing a $\ket{1}$ measurement outcome for the flag qubit. These errors and their syndromes are listed in Table \ref{table:flag_errors4} and Table \ref{table:flag_errors6}. Each of these errors has a distinct syndrome, so it can be corrected. Note that $Y$ errors on the ancilla qubit have the same effect on the data as $X$ errors.

As argued by Chao and Reichardt \cite{chao2017fault}, this method of syndrome measurement is fault-tolerant. Under the standard depolarizing model, all errors that appear with probability linear in the physical error rate can be detected and corrected using the above flag method, hence making the Bare $[[7,1,3]]$ code fault-tolerant under the standard depolarizing error model.\\

\begin{table}[h]
\begin{center}
\begin{tabular}{ c|c|c||c|c|c }
 \hline \hline
 b errors & Data error & syndrome & c errors & Data error & syndrome\\ \hline
 $IX$ & $Y_5 Y_6$ & $001101$ & $IX$ & $Y_6$ & $000100$\\
 \textcolor{red}{$XX$} & \textcolor{red}{$X_3 Y_5 Y_6$} & \textcolor{red}{$001111$} & \textcolor{red}{$XX$} & \textcolor{red}{$X_5 Y_6$} & \textcolor{red}{$000111$}\\
 \textcolor{red}{$YX$} & \textcolor{red}{$Y_3 Y_5 Y_6$} & \textcolor{red}{$001010$} & $YX$ & $Y_5 Y_6$ & $001101$\\
 $ZX$ & $Z_3 Y_5 Y_6$ & $001000$ & \textcolor{red}{$ZX$} & \textcolor{red}{$Z_5 Y_6$} & \textcolor{red}{$001110$}\\
\hline \hline
\end{tabular}
\captionsetup{justification=raggedright, singlelinecheck=false}
\caption{The nontrivial data errors that can result from a single two-qubit error for the stabilizer $Z_2 Z_3 Y_5 Y_6$ (Figure \ref{fig:weight4}) and trigger the flag qubit measurement. The errors marked in red can lead to logical errors with the bare method.}
\label{table:flag_errors4}
\end{center} 
\end{table} 

\begin{table}[h]
\begin{center}
\begin{tabular}{ c|c|c||c|c|c }
 \hline \hline
 b errors & Data error & syndrome & c errors & Data error & syndrome\\ \hline
 $IX$ & $Z_1 Z_2 Z_4 Z_5$ & $100010$ & \textcolor{red}{$IX$} & \textcolor{red}{$Z_1 Z_2 Z_5$} & \textcolor{red}{$010010$}\\
 $XX$ & $Z_1 Z_2 X_3 Z_4 Z_5$ & $100000$ & \textcolor{red}{$XX$} & \textcolor{red}{$Z_1 Z_2 X_4 Z_5$} & \textcolor{red}{$010011$}\\
 $YX$ & $Z_1 Z_2 Y_3 Z_4 Z_5$ & $100101$ & \textcolor{red}{$YX$} & \textcolor{red}{$Z_1 Z_2 Y_4 Z_5$} & \textcolor{red}{$100011$}\\
 $ZX$ & $Z_1 Z_2 Z_3 Z_4 Z_5$ & $100111$ & $ZX$ & $Z_1 Z_2 Z_4 Z_5$ & $100010$\\
\hline \hline
 d errors & Data error & syndrome & e errors & Data error & syndrome\\ \hline
 \textcolor{red}{$IX$} & \textcolor{red}{$Z_1 Z_5$} & \textcolor{red}{$011010$} & $IX$ & $Z_5$ & $001010$\\
 \textcolor{red}{$XX$} & \textcolor{red}{$Z_1 X_2 Z_5$} & \textcolor{red}{$011001$} & \textcolor{red}{$XX$} & \textcolor{red}{$X_1 Z_5$} & \textcolor{red}{$001011$}\\
 \textcolor{red}{$YX$} & \textcolor{red}{$Z_1 Y_2 Z_5$} & \textcolor{red}{$010001$} & \textcolor{red}{$YX$} & \textcolor{red}{$Y_1 Z_5$} & \textcolor{red}{$011011$}\\
 \textcolor{red}{$ZX$} & \textcolor{red}{$Z_1 Z_2 Z_5$} & \textcolor{red}{$010010$} & \textcolor{red}{$ZX$} & \textcolor{red}{$Z_1 Z_5$} & \textcolor{red}{$011010$}\\
\hline \hline
\end{tabular}
\captionsetup{justification=raggedright, singlelinecheck=false}
\caption{The nontrivial data errors that can result from a single two-qubit error for the stabilizer $Z_0 Z_1 Z_2 X_3 Z_4 Z_5$ (Figure \ref{fig:weight6}) and trigger the flag qubit measurement. The errors marked in red can lead to logical errors with the bare method.}
\label{table:flag_errors6}
\end{center} 
\end{table} 

\section{Simulation scheme}\label{sec:simu_scheme}

In this section we present the circuit used in the simulation, the QEC scheme used to perform correction after each round of the Bare $[[7,1,3]]$ code with noise, and the calculations we performed to obtain logical error rates. The simulations were done using the CHP stabilizer simulator \cite{CHP}. The simulation follows the 1-rectangle (1-Rec) formalism \cite{aliferis2005quantum} for distance-3 and level-1 encoding, where the 1-Ga we are simulating is the level-1 logical identity gate.

\subsection{Simulation Circuit}
\begin{figure*}
\centering
\includegraphics[width =\linewidth]{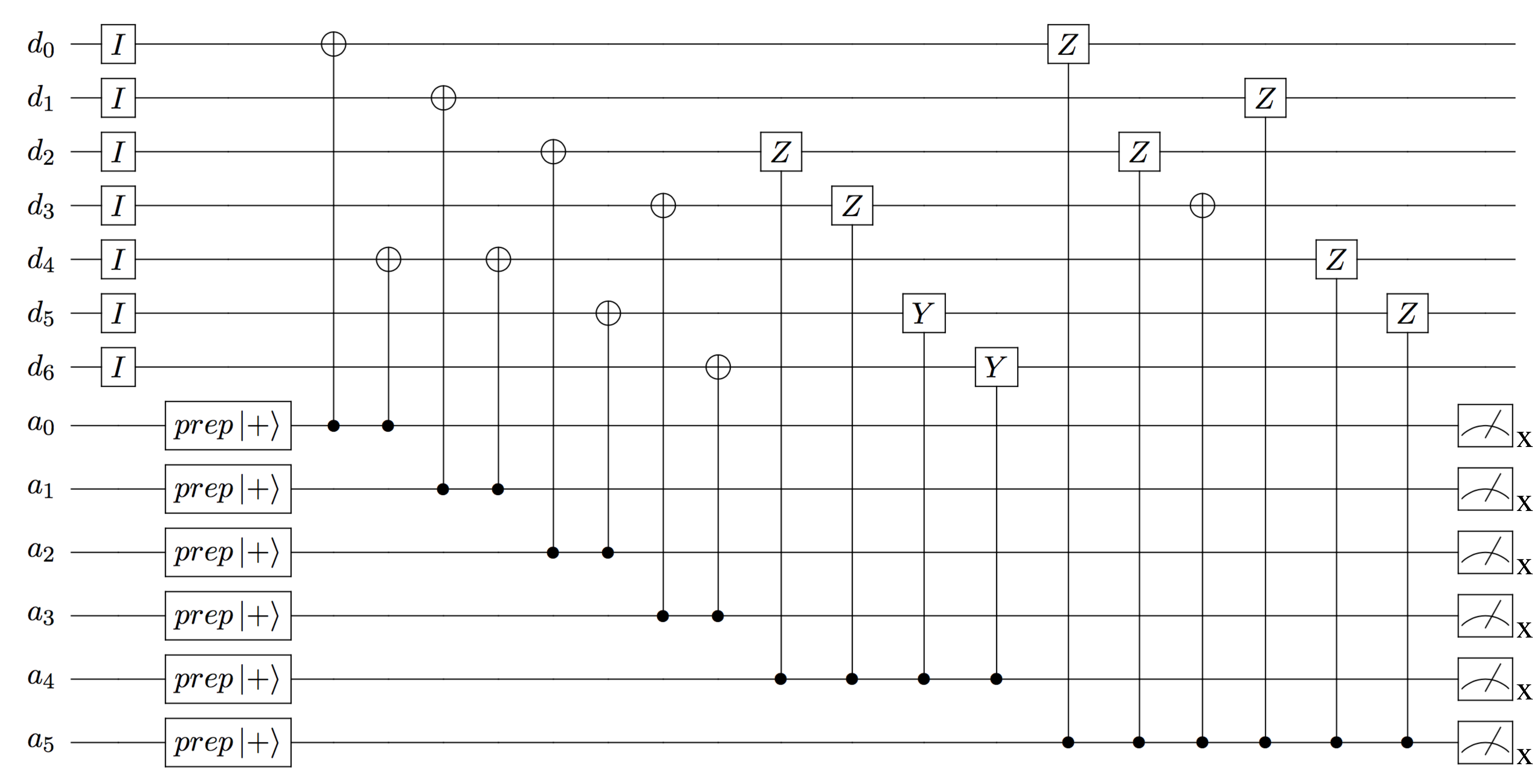}
\captionsetup{justification=raggedright, singlelinecheck=false}
\caption{The circuit simulating the Bare $[[7,1,3]]$ code and its measurement steps. The circuit is constructed with only Clifford gates, and errors (single-qubit and two-qubit) are inserted after ancillary qubit preparation, CNOT, Control-$Z$, Control-$Y$, and measurements.}
\label{fig:cross_qcircuit}
\end{figure*}

Fig. \ref{fig:cross_qcircuit} shows the quantum circuit we constructed to simulate one error-correcting step of the bare method for the Bare $[[7,1,3]]$ code. The simulated gate is a logical identity followed by an error-correction gadget. The gates are grouped together for each ancillary qubit measurement, and the order of qubit-coupling between the data qubits and ancillary qubits follows exactly the scheduling as required for fault-tolerant measurements of each stabilizer generators. For each gate shown in this circuit, errors of appropriate type are inserted after it with a probability given by the noise model. The circuit for simulating the flag method requires two additional flag qubits, and has a different order of qubit-coupling between the data and ancillary qubits, as described in Section \ref{sec:flag}.

The syndrome measurement is repeated up to three times with freshly prepared ancillary qubits.  The repetition is necessary due to errors during or between syndrome measurements \cite{macwilliams1977theory}. If the first two error syndromes agree, correction is performed based on the error syndrome.  If the two syndromes disagree, a third syndrome is measured and correction is performed based on the third syndrome.

\subsection{Logical Error Rate Calculation}

We calculate the logical error rate under the two different error models for various physical error strengths. For each run of in the Monte Carlo simulation, we initialize all the data qubits in state $\ket{0}$, 6 of the ancillary qubits in state $\ket{+}$ for syndrome measurements, and 2 of the ancillary qubits in state $\ket{0}$ to  be used as flag qubits for the high-weight stabilizers.  We then perform one round of noise-free stabilizer measurements to project the state of the data qubits to the logical $\ket{0}$ state.  Then the simulation proceeds as follows:
\begin{enumerate}
    \item Perform 2 or 3 rounds of error correction with random errors inserted using the importance sampling scheme.
    \item Apply the decoder to determine the corresponding error configuration and correct accordingly.
    \item Perform noise-free correction to the final state to project the state back to the codespace.
    \item If the final state is different from the initial one, count as one logical error.
\end{enumerate}

In the importance sampling scheme (see Appendix A.2), we classify the error configurations into subsets according to the number of single- and two-qubit errors present in the configuration.  The subset consisting of configurations of $s$ single-qubit errors and $t$ two-qubit errors is labelled by $(s,t)$.  For example, if during the execution of the QEC circuit, two single-qubit preparations were faulty and an error occurred after one of the CNOTs, this error configuration belongs to the subset $(2,1)$.  

We perform Monte Carlo simulations on selected error subsets and compute the logical error rate per subset, $p_L (s,t)$, by calculating the ratio of the successful runs over the total number of runs.  The total logical error rate, $p_L$, is calculated as a weighted average over the selected error subsets:
$$
p_L(p_s, p_t) = \sum _{s,t} A_{s,t} (p_s, p_t) p_L (s,t), 
$$
where $A_{s,t}$ is the probability of occurrence of subset $(s,t)$, that is, the total probability of occurrence of all error configurations with $s$ single-qubit errors and $t$ two-qubit errors.  

We use the calculated logical error rates at different physical error strengths to estimate the pseudothreshold of the code under a particular error model \cite{svore2005flow}. The pseudothreshold is the intersection between the physical error rate line $y = 2p/3$ and the logical error rate $p_L (p_s, p_t)$.  The reason for using this line instead of $y = p$ is that, if we assume a symmetric depolarizing noise model on a single qubit, the infidelity is given by $2p/3$.  More intuitively, if we focus on a single-qubit Pauli state, like $| 0 \rangle$, only two Pauli errors ($X$ and $Y$ in this case) will cause an error.  Furthermore, this will result in a slightly more pessimistic value for the pseudothreshold.

\section{Results}\label{sec:results}
In this section we present and analyze the results from numerical Monte Carlo simulations of the Bare $[[7,1,3]]$ code under different noise models using the two different syndrome measurement methods. We first focus on the standard depolarizing error model and then on the anisotropic error model.  We show that using the bare method, the Bare $[[7,1,3]]$ code does not have a pseudothreshold under the standard depolarizing error model, due to some uncorrectable two-qubit errors that have probability linear in the physical error strength. 

\subsection{Standard Depolarizing Error Model}

\begin{figure}[h]
\centering
\includegraphics[width =\linewidth]{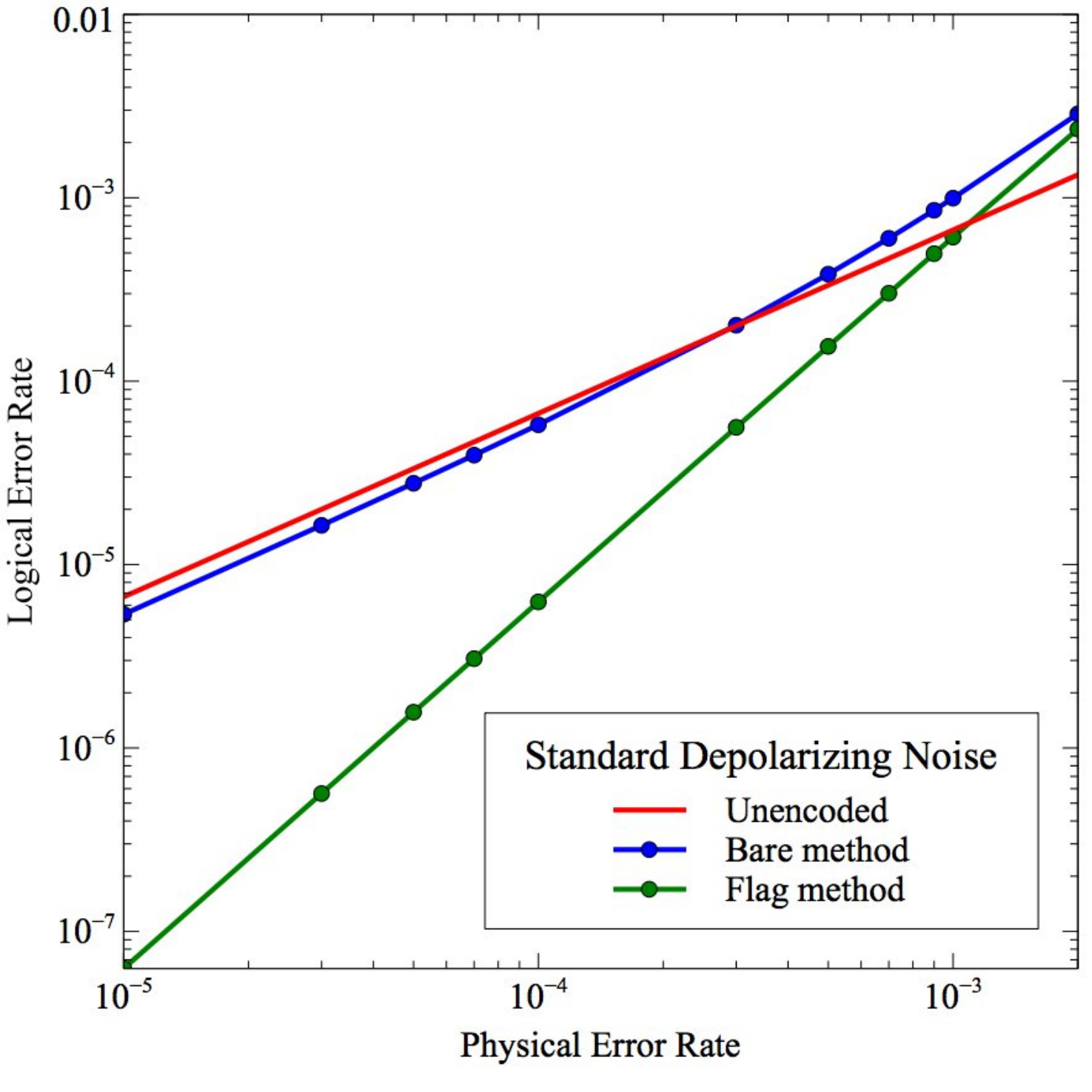}
\captionsetup{justification=raggedright, singlelinecheck=false}
\caption{Logical error rate for the Bare $[[7,1,3]]$ code under the standard error model. For the bare method, the logical error rate of the code remains parallel to the physical line for physical error rates below the intersection.  This implies that the logical error rate is linear in $p$.   For the flag method, we observe a level-1 pseudothreshold of $1.08 \times 10^{-3}$.}
\label{fig:standard}
\end{figure}

Fig. \ref{fig:standard} presents the logical error rates for the standard depolarizing model with $p_s = p_t = p$ for several values of $p$.  For the bare method, the best fit corresponds to a function with linear term as leading term, while for the flag method the best fit corresponds to a function with quadratic term as leading term.

For the bare method, although there is an intersection between the logical error rate and the physical error rate, the curves still remain parallel for physical error rates below the intersection.  The fact that there is still a linear term in the logical error rate implies that we would not observe an exponential suppression on the logical error rate with subsequent levels of concatenation.  Thus, this intersection is not a pseudothreshold for the Bare $[[7,1,3]]$ code, and the code is not fault-tolerant under the standard depolarizing error model. Examples of why this would happen are shown in Fig. \ref{fig:error}.

For the flag method, all errors that occur with probability linearly proportional to the physical error rate can be corrected.  By performing a quadratic fit for the logical error rates and computing the intersection of the fitting curve with the error rate of an unencoded qubit, we obtain a level-1 pseudothreshold of $1.08 \times 10^{-3}$.

\subsection{Anisotropic Error Model}
We now calculate the pseudothreshold of the Bare $[[7,1,3]]$ code under the anisotropic error model. In this model, the physical error strength after each gate is $p$, and $p_s = p_t = p$.

\begin{figure}[ht]
\centering
\includegraphics[width =\linewidth]{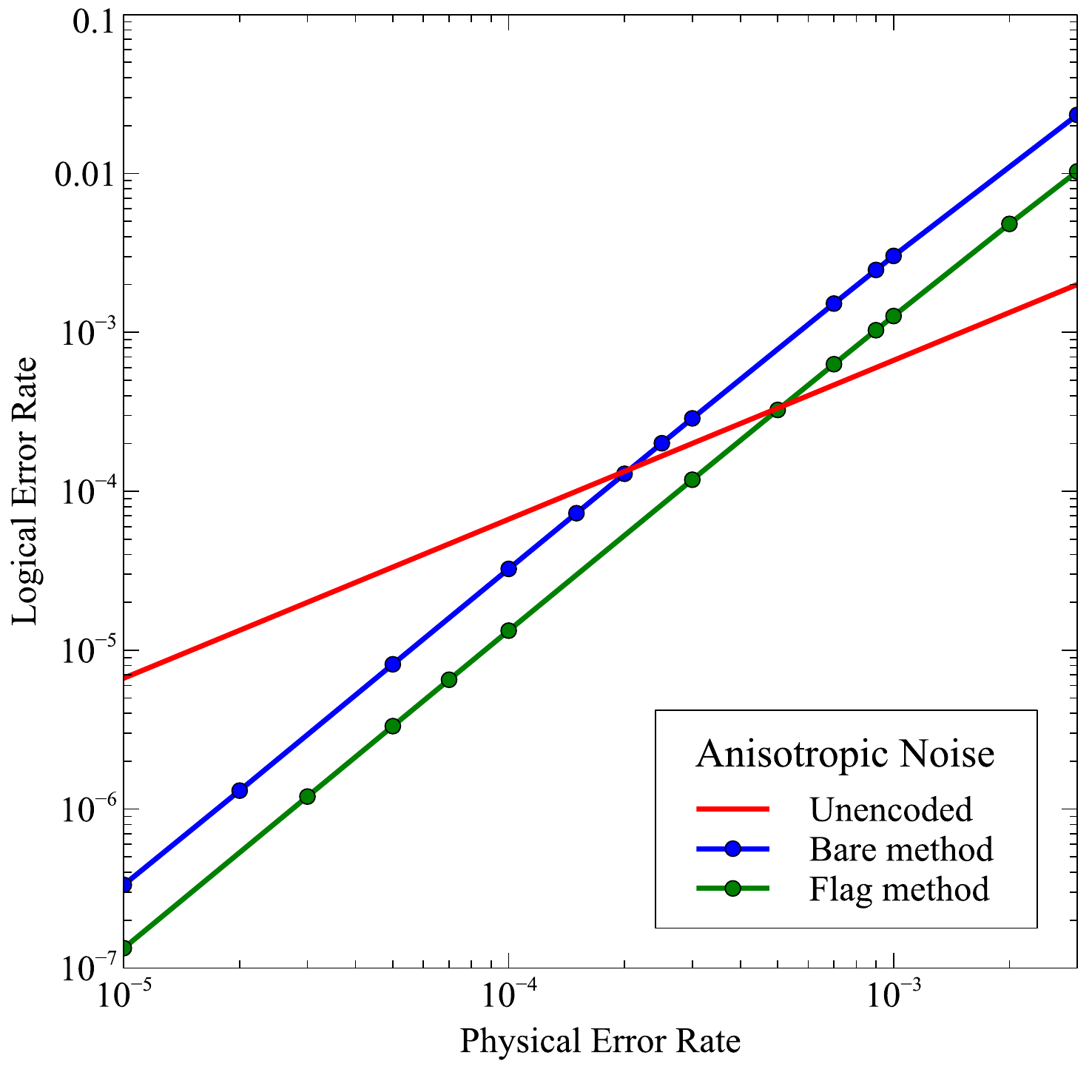}
\captionsetup{justification=raggedright, singlelinecheck=false}
\caption{Logical error rate for the Bare $[[7,1,3]]$ code under the anisotropic error model, $p_{s} = p_{t}$. For the bare method, the level-1 pseudothreshold is at $p=2.0 \times 10^{-4}$. For the flag method, the level-1 pseudothreshold is at $p=4.0 \times 10^{-4}$. The Bare $[[7,1,3]]$ code is fault-tolerant under the anisotropic error model with both syndrome measurement methods.}
\label{fig:iontrap}
\end{figure}

By performing a quadratic fit for the logical error rates and computing the intersection of the fitting curve with the error rates of an unencoded qubit, we obtain level-1 pseudothreshold for the Bare $[[7,1,3]]$ code with both syndrome measurement methods. This is because all single-qubit and two-qubit errors that occur with probability linear in $p$ are guaranteed to be detected and corrected by the Bare $[[7,1,3]]$ code.  From Fig. \ref{fig:iontrap} we can see that for the bare method, the level-1 pseudothreshold is at $2.0 \times 10^{-4}$. For the flag method, the level-1 pseudothreshold is at $4.0 \times 10^{-4}$.

The main reason for the difference between the two pseudothresholds is that the number of uncorrectable error configurations with probability quadratically proportional to the physical error rate is higher for the bare method.  Consider, for example, that some errors with probability quadratically proportional to the physical error rate can behave the same as the two-qubit errors listed in Table \ref{table:flag_errors4} and Table \ref{table:flag_errors6}. Those errors could lead to logical errors with the bare method, but can be corrected with the flag method.

\subsection{Comparison of Results across Error Models}
From the results presented in Fig. \ref{fig:standard} and Fig. \ref{fig:iontrap} we can see that the level-1 pseudothreshold under the anisotropic error model is lower than that under the standard depolarizing error model. This is because in the anisotropic error model, after each two-qubit gate we are simulating a single-qubit random Pauli error on each qubit with probability $p_s$ (Section \ref{subsec:error}), leading to more possible error locations in the overall circuit. If we assume the single-qubit errors and two-qubit errors are equally probable, as we did in the results shown in Fig. \ref{fig:standard} and Fig. \ref{fig:iontrap}, errors after the two-qubit gates are 3 times more likely to occur under the anisotropic error model than under the standard depolarizing model, because of the extra single-qubit Pauli errors.  Hence, the Bare $[[7,1,3]]$ code under the anisotropic error model has a lower pseudothreshold than under the standard depolarizing model.

\section{Conclusions} \label{sec:conclusions} 

In this paper we have presented properties of a new $[[7,1,3]]$ stabilizer code that can achieve fault-tolerant syndrome measurements using a single ancillary qubit under the anisotropic error model, and showed why certain two-qubit errors under the standard depolarizing error model would prevent this code from achieving fault-tolerant measurements. In particular, the limit on total number of syndrome outcomes makes it impossible for the lookup table decoder to detect and correct all two-qubit errors in the standard depolarizing error model. Additionally, we showed that using the flag method the Bare $[[7,1,3]]$ code can achieve fault-tolerance under both the standard depolarizing model and the anisotropic model. Overall, with the flag method the Bare $[[7,1,3]]$ code shows better performance under realistic error models. 

We note that the $[[7,1,3]]$ triangle code presented by Yoder and Kim  \cite{yoder2016surface} also does not require additional ancillary qubit preparations and is able to achieve fault-tolerance under the standard depolarizing error model. However, their code would not be identified by the search criteria that identified our code \cite{private}: a syndrome measurement with strictly bare ancillary qubits on their code would not be able to correct all hook errors originated from a single qubit error on the ancillary qubit. Instead they proposed an interwoven syndrome measurement method to achieve fault-tolerance. Since the interwoven method is not applicable for the Bare $[[7,1,3]]$ code, no direct comparison between the two codes can be easily made. But if the flag method \cite{chao2017fault} was to be applied on both codes, they can both achieve fault-tolerance under the standard depolarizing error model with the same amount of resources, and the pseudothreshold we obtained for the Bare $[[7,1,3]]$ code for the 1-Rec is comparable to the pseudothreshold reported by Yoder and Kim for the exREC \cite{aliferis2005quantum}.

A natural direction of future work would be to study the behavior of the Bare $[[7,1,3]]$ code under realistic other error models such as coherent errors and amplitude damping \cite{gutierrez2013approx}, and tailor this code for architecture-specific noise environments where limitations on ancillary qubit resources and measurement times are required.

The fact that the Bare $[[7,1,3]]$ code with the bare method does not perform well under the theoretical standard error model but can achieve fault-tolerant measurements under a more realistic error model suggests that more effort should be put into designing and finding error correcting codes that help achieve fault-tolerant quantum computation under realistic scenarios \cite{flammia2017tailored}. As realistic error models are becoming more and more relevant, development of environment-specific error correcting codes will become more important.

\begin{acknowledgments}
This work would not have been possible without the help of Andrew Cross. We would like to thank him for discovering this code, for encouraging us to study its properties, and for discussions and comments on the manuscript. We thank Colin Trout and Theodore Yoder for comments on the manuscript and discussions on the flag method. This work was supported by the Office of the Director of National Intelligence - Intelligence Advanced Research Projects Activity through ARO contract W911NF-10-1-0231 and the National Science Foundation grant PHY-1415461.

\end{acknowledgments}

\appendix
\section{Efficient Sampling Algorithm for Monte Carlo Simulations}\label{sec:MC_algos}

For a given QEC circuit and a physical noise model, obtaining an exact algebraic expression for the logical error rate is in principle possible.  In the context of a stabilizer code and a noise model consisting of stochastic discrete errors, this amounts to ($1$) enumerating every possible error configuration on the circuit, ($2$) calculating its probability of occurrence, and ($3$) determining whether or nor it results in a logical error.  The logical error rate is then given by:
\begin{equation}
    p_L = \sum_{i=0} ^{N_c} A_i \, p_i,
\end{equation}
where $N_c$ is the total number of error configurations, $A_i$ is the probability of occurrence of error configuration $i$, and $p_i = 0 (1)$ if the error configuration $i$ is correctable (uncorrectable).  Although possible in principle, the exact computation of a logical error rate is infeasible in practice, due to the high cardinality of the error configuration set.  For a circuit with $n_g$ gates and $s$ possible different errors after each gate, $N_c = (s+1)^{n_g}$.
Even a modest circuit, like a level-1 QEC routine for a distance-$3$ code consisting of $3$ rounds of stabilizer measurements, contains more than $100$ gates, making the exact computation of the logical error rate impractical.  

It is, therefore, common to employ Monte Carlo methods to estimate $p_L$ \cite{knill2005noisydevices, raussendorf2007fault, fowler2009high, smitha2016FPT}.   The basic procedure to obtain a logical error rate consists of two steps: ($1$) the generation of a faulty circuit (an error configuration) based on the physical noise model and ($2$) the simulation of the circuit to determine if that particular error configuration is correctable.  In this section, we describe the two different Monte Carlo methods used in the first step.  The first one utilizes traditional direct sampling of the whole error configuration set.  The second one relies on importance sampling of the error configuration subsets that are relevant to the logical error rate.  Both methods are based on direct sampling, i.e., each error configuration is completely uncorrelated from the previous ones.

\subsection{Traditional Sampler}

To generate an error configuration, the traditional sampler traverses the circuit exhaustively and after each gate an error is inserted with a probability given by the noise model.  This approach is convenient for high error rates.  However, it is problematic if the error rate is low, because most of the times no error is inserted.  For illustrative purposes, consider a simple case where the error rate $p = 0.1\%$ is the same for ech gate and the circuit has $N_g = 100$ gates.  The traditional sampler will return an error-free circuit $(1-p)^{n_g} \approx 90\%$ of the runs.  Furthermore, if the circuit corresponds to a fault-tolerant QEC protocol of a distance-$3$ code, by construction no error configuration of weight-$1$ will result in a logical error.  This implies that $(1-p)^{n_g} + n_g p (1-p)^{n_g - 1} \approx 99.5\%$ of the runs will generate an error configuration which is known a priori to be correctable.  The limitation of the traditional sampler becomes even more dramatic for lower error rates and codes of higher distance \cite{bravyi2013rare}.  

\subsection{Importance Sampler}

It is possible to split the error configuration set into subsets based on each configuration's error weight (number of errors).  The key advantage of the importance sampler relies on two features of this particular subset splitting: ($1$) it is straightforward to compute analytically the total probability of occurrence of each subset (the sum of the probabilities of occurrence of the error configurations in the subset) and ($2$) for low error rates we can safely ignore high weight error subsets, since their probability of occurrence will be vanishingly small.   

Here we consider subsets in terms of both errors that occur after single-qubit gates and two-qubit gates.  For example, a subset labelled by $(s, t)$ contains all error configurations of the circuit with $s$ errors after single-qubit gates and $t$ errors after two-qubit gates.  Notice that an error after a two-qubit gate can still be of weight $1$.  Let $n_s$ and $n_t$ be the total number of single-qubit gates and two-qubit gates in this circuit.  We assume an error model where the error probability is the same for all single-qubit gates ($p_s$) and the error probability is also the same for all two-qubit gates ($p_t$).  The probability of occurrence of the error subset $(s,t)$ is then:
\begin{equation} \label{eq:subset_weight}
    A_{s,t}(p_s, p_t) = \dbinom{n_s}{s}p_s^s (1-p_s)^{n_s-s} \dbinom{n_t}{t}p_t^t (1-p_t)^{n_t-t}.
\end{equation}

Given a quantum circuit and a noise model, the algorithm to estimate the logical error rate using the importance sampler consists of three steps: 

\begin{enumerate}
    \item Select a tolerance value.  This corresponds to the total added probability of occurrence of the high weight subsets that will not be sampled.  In the worst scenario imaginable, every error configuration in the excluded subsets would result in a logical error.  The tolerance value represents the worst-case discrepancy between the real and the obtained logical error rate.  In particular, it provides a worst-case upper bound to the logical error rate.     
    \item For each error subset $(s,t)$, use direct Monte Carlo sampling to approximate its logical error rate $p_L (s,t)$.  The sampling is done by randomly selecting $s$ single-qubit gates and $t$ two-qubit gates and adding errors after them.  
    \item Calculate the total logical error rate for the circuit:
    \begin{equation} \label{eq:total_pl}
        p_L(p_s, p_t) = \sum_{s,t} A_{s,t}(p_s, p_t) p_L (s,t),
    \end{equation}
    where $A_{s,t}(p_s,p_t)$ is computed analytically using Equation \ref{eq:subset_weight}.
\end{enumerate}

Notice that the logical error rate $p_L (s,t)$ for a particular $(s,t)$ subset is not a function of the physical error rates $p_s$ and $p_t$.  This means that, for a given circuit and noise model, we can pre-compute the logical error rate for each relevant subset, and use those values to evaluate the total logical error rate for different physical error rates.  In contrast to the traditional sampler, there is no need to re-run the Monte Carlo simulations for different values of $p_s$ and $p_t$.  We simply calculate new values for the probabilities of subset occurrence $A_{s,t}$ and compute $p_L$ using Equation \ref{eq:total_pl}.  Once the Monte Carlo simulations for each relevant subset are done, the logical error rates for different physical error rates can be computed at no additional cost, making the importance sampler much more efficient than the traditional one.  However, for high physical error rates (about $10^{-2}$ and higher) the importance sampler becomes either inaccurate or very slow.  The trade-off arises because as the error rate increases so does the probability of occurrence of higher weight subsets and therefore the number of subsets that need to be included to obtain an accurate result.  In this error regime, it is suitable to employ the traditional sampler.

\subsection{Comparing Performance of Sampling Algorithms }
\begin{figure*}
    \centering
    \begin{subfigure}[b]{0.45\textwidth}
    \includegraphics[width=\linewidth]{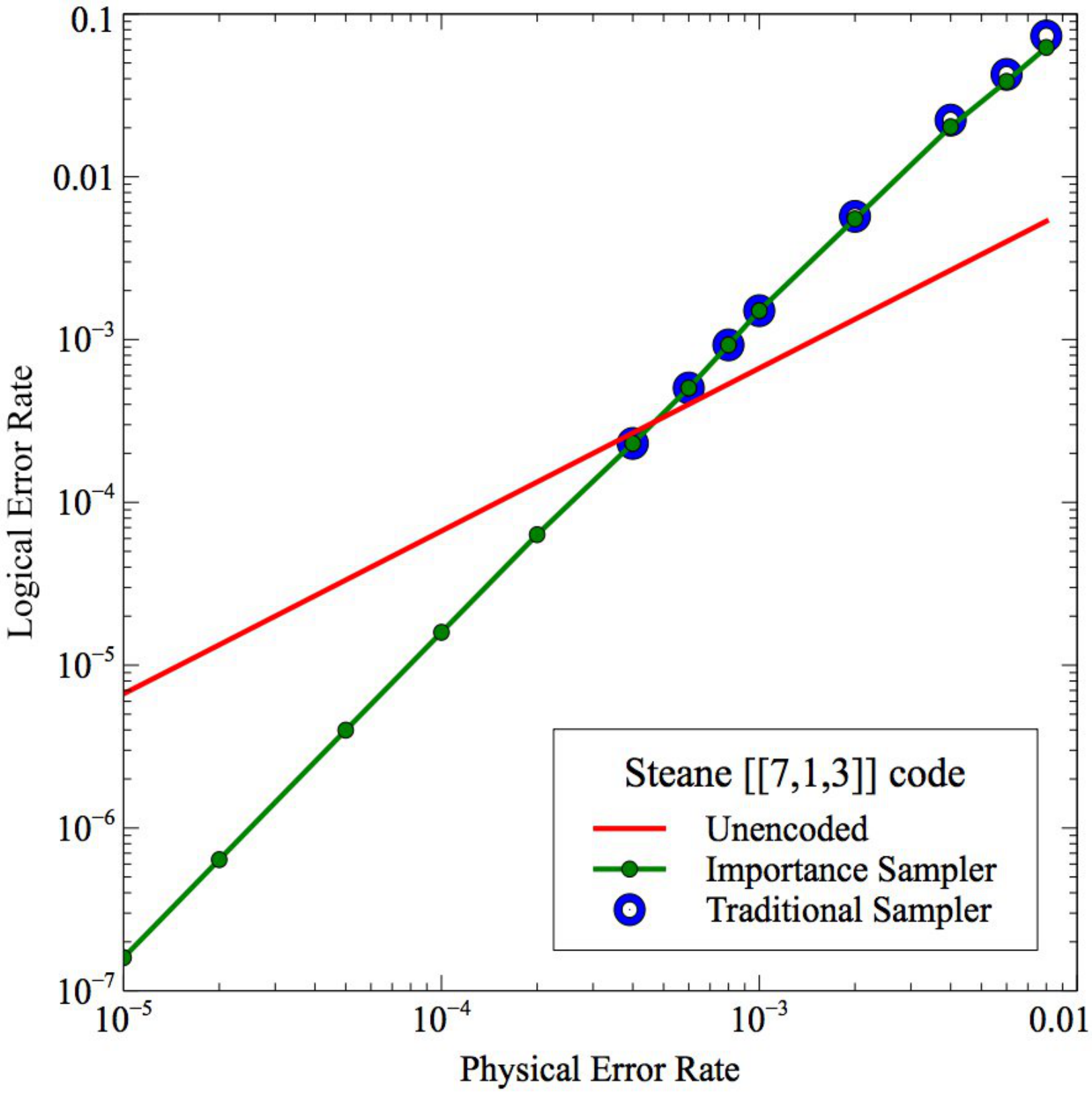}
    \caption{Steane $[[7,1,3]]$ code}
    \label{fig:steane_standard}
    \end{subfigure}
    ~ 
    \begin{subfigure}[b]{0.45\textwidth}
    \includegraphics[width = \linewidth]{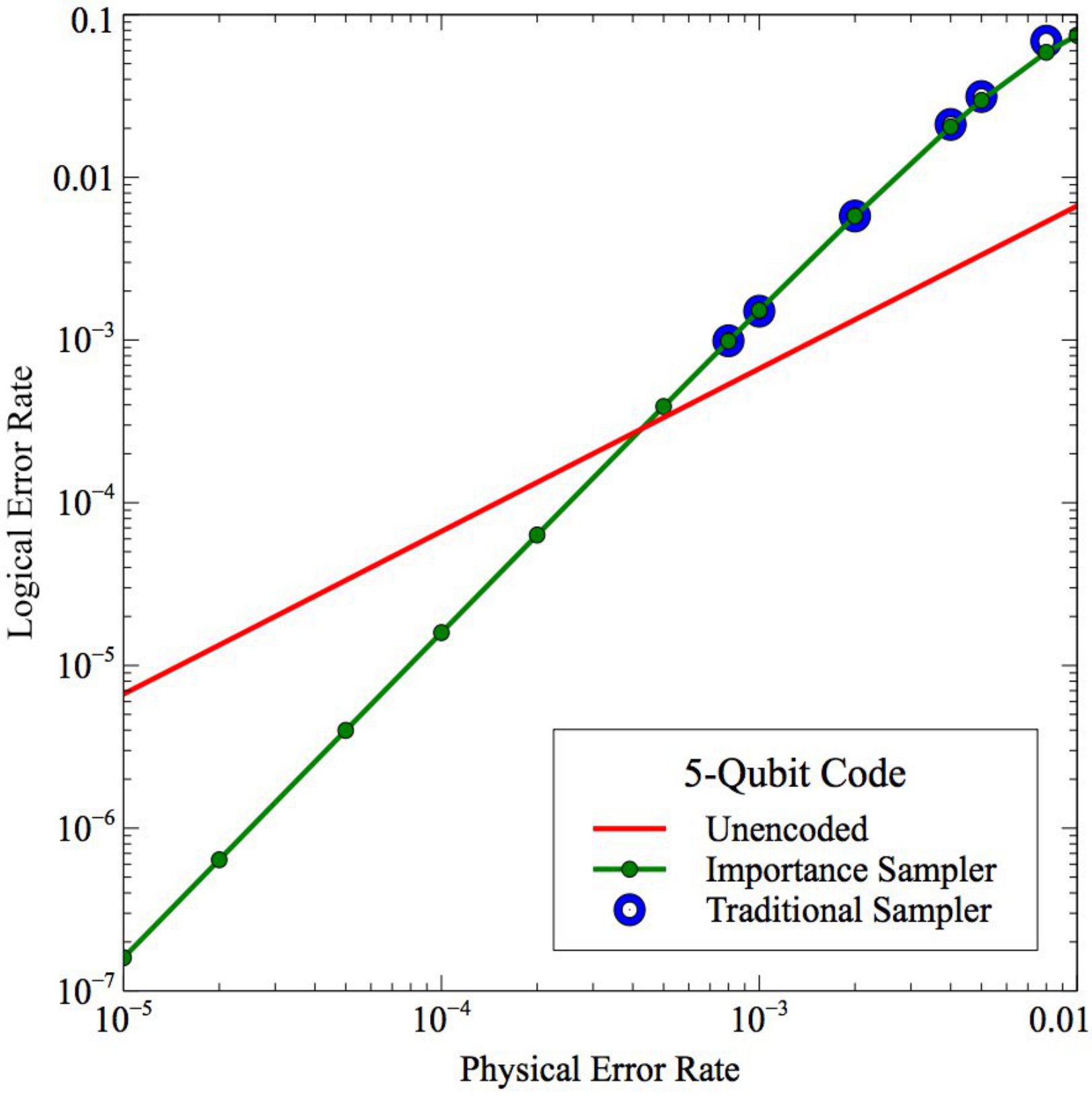}
    \caption{Five-qubit code}
    \label{fig:fivequbit_standard}
    \end{subfigure}
    \captionsetup{justification=raggedright, singlelinecheck=false}
    \caption{Logical error rate with the importance sampler and the traditional sampler for the Steane $[[7,1,3]]$ code and the five-qubit code with Shor-style ancillary qubits under the standard depolarizing error model.  Up to about $p = 2.0 \times 10^{-3}$, the two samplers result in essentially the same logical error rate.}
    \label{fig:test_standard}
\end{figure*}

\begin{figure*}
    \centering
    \begin{subfigure}[b]{0.45\textwidth}
    \includegraphics[width=\linewidth]{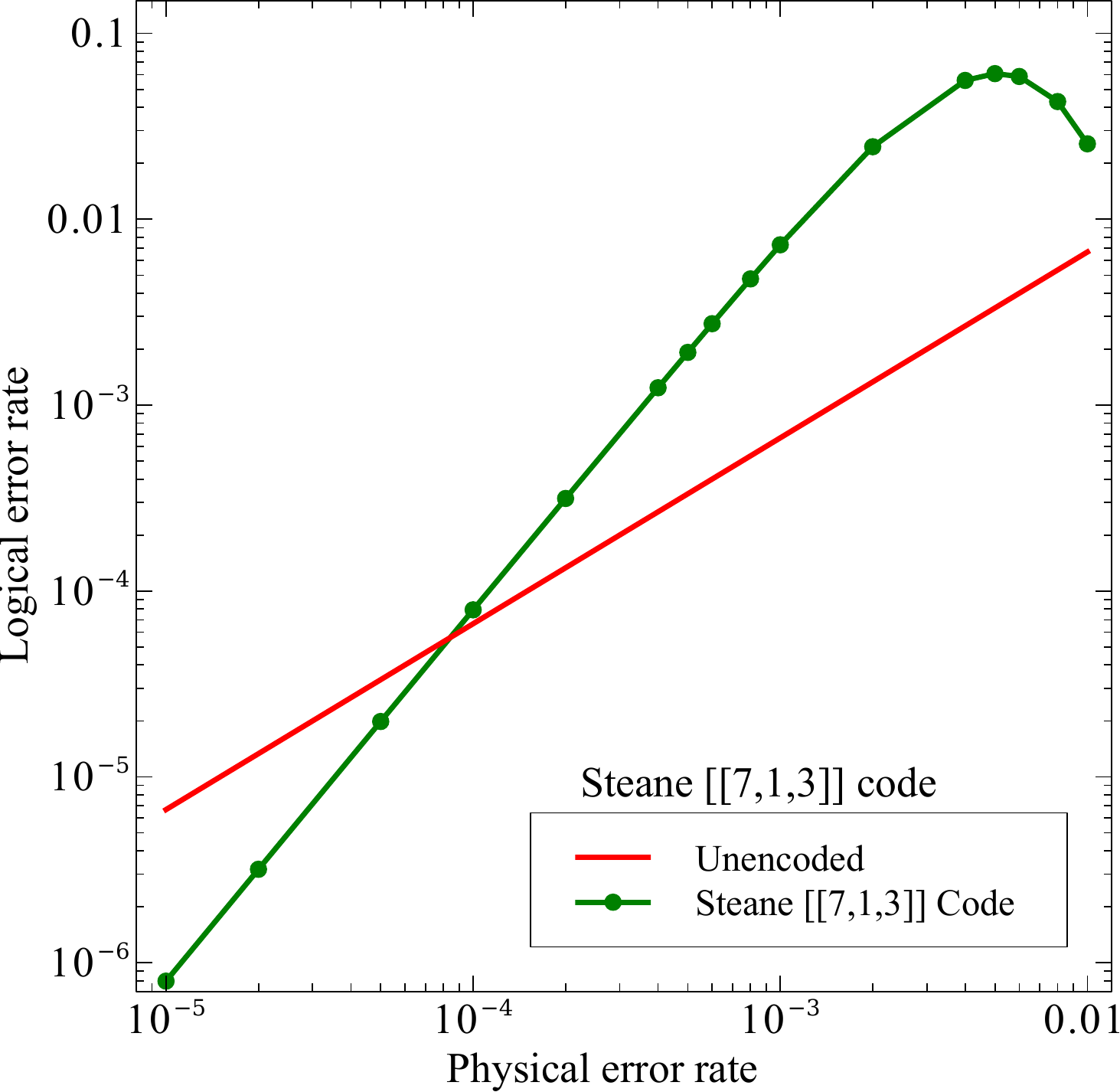}
    \caption{Steane $[[7,1,3]]$ code}
    \label{fig:steane_ion}
    \end{subfigure}
    ~ 
    \begin{subfigure}[b]{0.45\textwidth}
    \includegraphics[width = \linewidth]{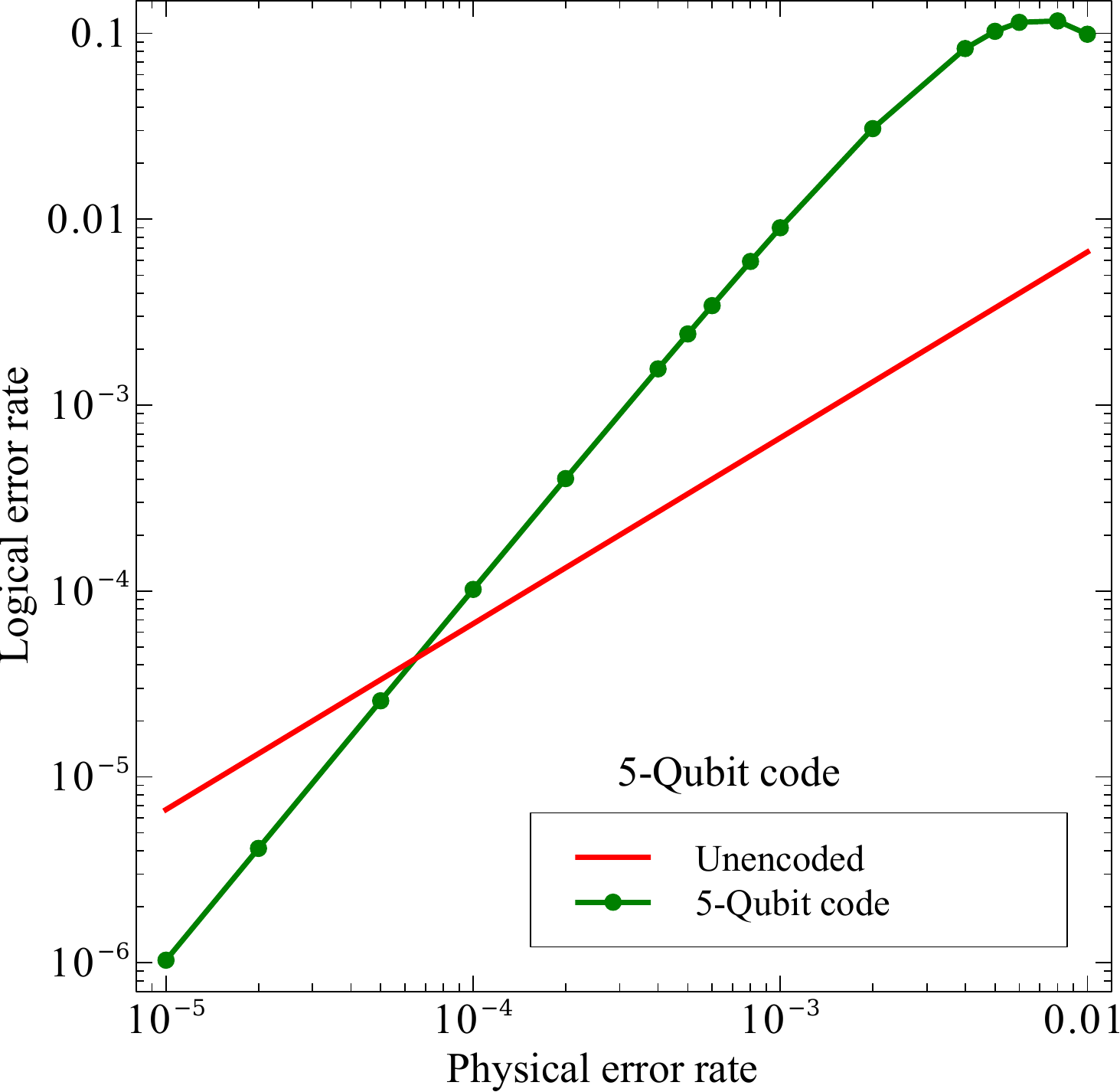}
    \caption{Five-qubit code}
    \label{fig:fivequbit_ion}
    \end{subfigure}
    \captionsetup{justification=raggedright, singlelinecheck=false}
    \caption{Logical error rate with the importance sampler for the Steane $[[7,1,3]]$ code  and the five qubit code with Shor-style ancillary qubits under the anisotropic error model.  Around $p = 1.0 \times 10^{-2}$, the probability of occurrence of the high-weight errors becomes significant and ignoring them causes a dip on the logical error rate.  The dip occurs at a lower $p$ under the anisotropic error model because there are more faulty locations than under the standard model.}
    \label{fig:test_ion}
\end{figure*}

In order to test the performance of the importance sampler, we simulated the logical error rate for the Steane $[[7,1,3]]$ code, the five-qubit code, and the Bare $[[7,1,3]]$ code with both the traditional and the importance sampling algorithm. To make sure that the results of the two sampling algorithms are comparable, we analytically proved that the probability for an arbitrary error to occur at any location in the circuit is the same for both sampling algorithms.\\
From both Fig. \ref{fig:test_standard} and Fig. \ref{fig:test_ion} we can see that the results of the two sampling algorithms coincide exactly at lower physical error rates, but can start to diverge as physical error rate increases, with the logical error rate obtained from the traditional sampler slightly higher than that from the importance sampler. This is because for a given number of subsets, at lower physical error rates these subsets are enough to calculate the cumulative logical error rate to a high accuracy, but for higher physical error rates the importance sampler must sample a larger number of error subsets in order to achieve the same level of accuracy in the cumulative logical error rate. In this case, the traditional sampler accounts for all error subsets of this circuit, while the importance sampler only accounts for a small number of subsets, resulting in a smaller logical error rate.

\newpage
\bibliography{References}
\bibliographystyle{apsrev}

\end{document}